\documentclass[11pt]{article}
\usepackage{pslatex}
\usepackage{amsmath}
\usepackage{amssymb}
\usepackage{bbm}
\usepackage{euscript}
\usepackage{epsfig}
\usepackage{graphicx}
\usepackage{setspace}
\usepackage{cite}
\usepackage[colorinlistoftodos,prependcaption,textsize=tiny]{todonotes}
\usepackage{hyperref}

\newcommand{\pd}{\partial}
\renewcommand{\d}[1]{\mathrm{d}{#1}\,}

\usepackage{ifthen}
\usepackage{amsmath}
\usepackage{amssymb}
\usepackage{mathtools, nccmath}
\usepackage{graphicx, wrapfig}
\usepackage{slashed, bm}
\usepackage[utf8]{inputenc}

\usepackage{authblk}
\usepackage[normalem]{ulem}

\newcommand{\kt}{{\bm k}}

\newcommand{\qt}{{\bm q}}
\newcommand{\Dt}{{\bm \Delta}}

\definecolor{kkcolor}{rgb}{0,0.6,0.4}

\definecolor{mhcolor}{rgb}{0.7,.1,0.7}

 \title{Maximally entangled proton and charged hadron multiplicity
 in Deep Inelastic Scattering}

\author[1]{Martin~Hentschinski}
\author[2,3]{Krzysztof~Kutak}
\author[4]{Robert~Straka}

\affil[1]{\normalsize Departamento de Actuaria, F\'isica y Matem\'aticas, 
Universidad de las Americas Puebla, San Andr\'es Cholula, 72820 Puebla, Mexico }
\affil[2]{ \normalsize Institute of Nuclear Physics, Polish Academy of Sciences, 
 ul.~Radzikowskiego 152, 31-342, Krak\'ow, Poland}
 \affil[3]{ Brookhaven National Laboratory, Physics Department, Bldg. 510A, 20 Pennsylvania Street, 30-059 Upton, NY 11973 USA
 }
 \affil[4]{ \normalsize AGH University of Science and Technology, Krak\'ow, Poland}
 
\begin{document}

\maketitle
\begin{abstract}
We study the proposal by Kharzeev-Levin  to determine entanglement entropy in Deep Inelastic Scattering (DIS) from parton distribution functions  (PDFs) and to relate the former to the entropy of final state hadrons. We find several uncertainties in the current comparison to data, in particular the overall normalization, the relation between charged versus total hadron multiplicity in the comparison to experimental results as well as different methods to determine the number of partons in Deep Inelastic Scattering. We further provide a comparison to data based on leading order HERA PDF as well as PDFs obtained from an unintegrated gluon distribution subject to next-to-leading order Balitsky-Fadin-Kuraev-Lipatov  and Balitsky-Kovchegov evolution. Within uncertainties we find good agreement with H1 data. We provide also predictions for entropy  at lower photon virtualities, where  non-linear QCD dynamics is expected to become relevant.
\end{abstract}

\section{Introduction}
\label{sec:entanglement-entropy}

Entanglement is a nonlocal correlation unique to quantum systems \cite{Horodecki:2009zz}, see also the reviews \cite{Casini:2022rlv,Headrick:2019eth}. 
There are various proposals how to study entanglement in high energy physics such as  neutrino oscillations, spin correlations of $t-\bar t$ quarks or $\Lambda$ hyperons \cite{Kayser:2010bj,Afik:2020onf,Fabbrichesi:2021npl,Gong:2021bcp}. A measure of entanglement which is of particular interest is entanglement entropy \cite{Kovner:2015hga}. 
In \cite{Kharzeev:2017qzs} it has been proposed that one can associate  entanglement entropy with
the system of partons probed in Deep Inelastic Scattering (DIS) experiments. The proposal necessarily requires a measurement process which introduces a bi-partition to the system. The measured system is no longer in a pure quantum state and as a consequence entropy arises. The proposal in \cite{Kharzeev:2017qzs}
considers coordinate space entropy and is motivated by exact results obtained in conformal field theories in 1+1 dimensions \cite{Holzhey:1994we,Calabrese:2004eu}. The 1+1 dimensional picture provides reasonable guidance, since the basic quantity which describes the system of partons at some resolution scale is the integrated density of partons, {\it i.e.} the parton distribution function (PDF). It provides information about the one dimensional spatial structure of protons, although the underlying dynamics may take place in transverse dimension as well. 
For  further developments and other approaches see \cite{Kutak:2011rb,Peschanski:2012cw,Stoffers:2012mn,Kovner:2018rbf, Armesto:2019mna,Duan:2020jkz,Dvali:2021ooc,Ramos:2020kaj,Kharzeev:2021nzh,Hagiwara:2017uaz,Tu:2019ouv,Zhang:2021hra,Neill:2018uqw,Liu:2022qqf,Liu:2022hto,Liu:2022ohy,Dumitru:2022tud, Ramos:2022gia,
Duan:2021clk}. \\

The measurement process is provided by a probe. In DIS this probe is given by the virtual photon that is exchanged between electron and proton. Since the photon probes  only parts of the partonic system of the
proton, the remaining part has to be integrated out or traced over, giving rise to a reduced density matrix and therefore entanglement entropy. We note that this approach was also used to estimate entanglement entropy produced in proton-proton collisions at the LHC \cite{Tu:2019ouv}. It can be motivated  as follows:
for DIS at low $x$ and referring to the proton rest frame, it is possible to take color dipoles as the basic degrees of freedom. Color dipoles are color singlets and therefore natural candidates to generate entanglement entropy\footnote{See also \cite{Liu:2022hto} where the dipole degrees of freedom are used to obtain entanglement entropy}. The mechanism of the entanglement is the following:
\begin{figure}
    \centering
    \includegraphics[width=0.79\textwidth]{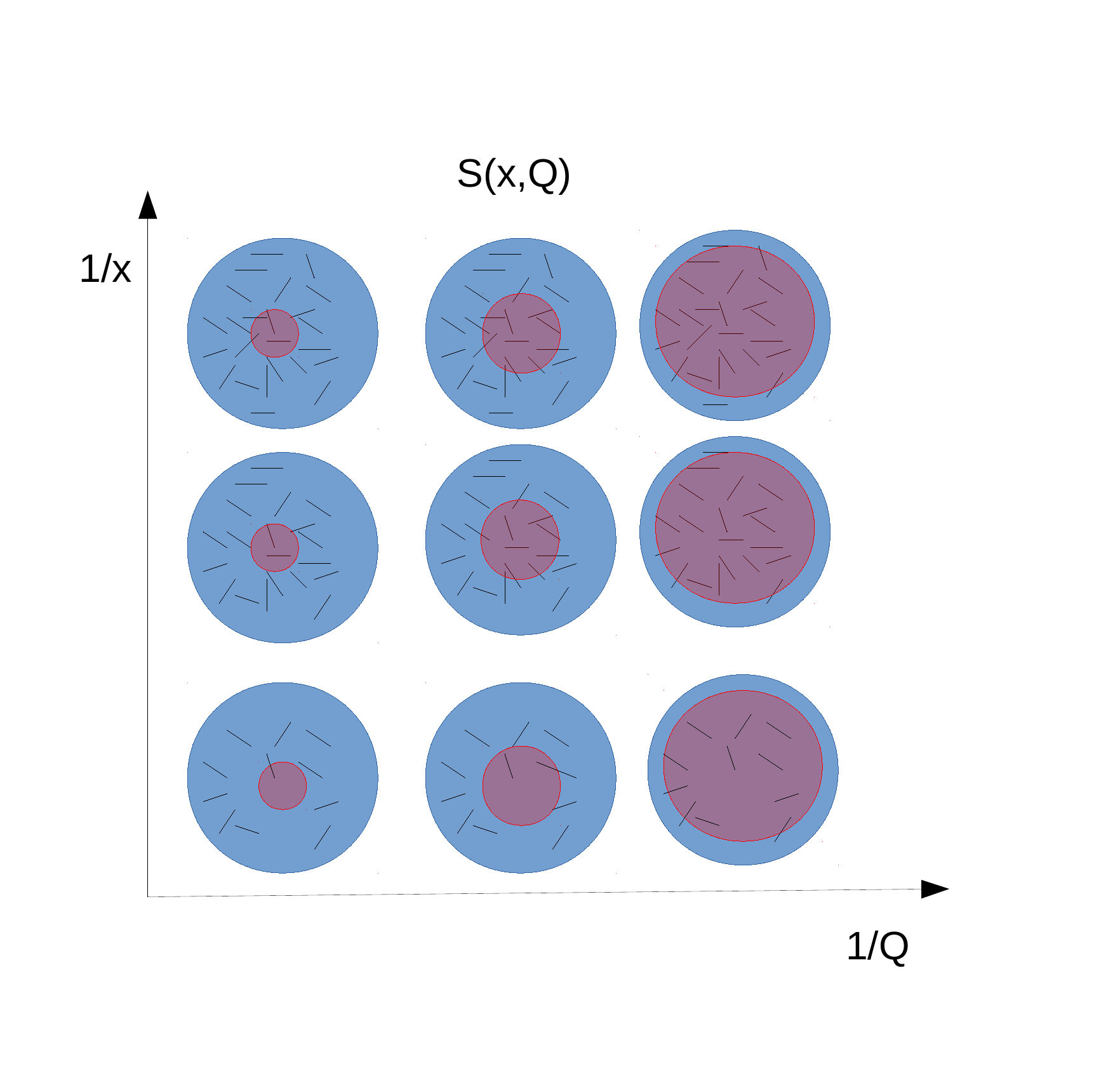}
    \caption{The figure illustrates the generation of entanglement in DIS within the dipole picture with $Q$ the photon virtuality. The blue region represents the proton, the red circle the area singled out by the virtual photon. The segments represent color dipoles in color singlet states. The entanglement  arises due to dipoles  that are  partially in the red and blue region.}
    \label{fig:entanglementfig}
\end{figure}
\begin{itemize}
    \item once the virtual photon resolves substructure in the proton, it singles out a certain region of area $\sim 1/Q^2$ see Fig.~\ref{fig:entanglementfig}.
    \item it might happen that one of the quarks that is a constituent of the  dipole is inside the  singled  out region  A  and the other part is in the region  B. Since the dipole is  a color  singlet state, the quark-antiquark system is strongly correlated and provides the source of entanglement between both regions. Due to the interaction, the dipole breaks up and the quark in the region B gives rise to the observed final state hadrons.
    \item the other quark gives a contribution to the  parton distribution function measured in the DIS process
    \item since both constituents of the dipole were in different regions, the entropy of both regions is identical. Only dipoles that bridge the regions give rise to the observed entropy.
    \item as one goes to higher energies the number of dipoles increases (higher twists)
    and more and more dipoles are partially in region A and region B. As a consequence, entropy grows.
    \item turning to smaller $Q^2$,  more and  more dipoles are completely inside region A and the entropy should eventually decrease.
\end{itemize}
The framework presented in  \cite{Kharzeev:2017qzs} provides furthermore an explicit formula on how to calculate this entanglement entropy as well as means how to obtain the latter from data through the determination of the entropy of the observed final state hadrons. Since the entire approach is based on distributions of color dipoles, which themselves can be derived from QCD within high energy factorization with $x\ll 1$ the expansion parameter, the proposed description is naturally restricted to the low $x$ regime. NLO fits of low $x$ gluon distribution are found to provide a good description of HERA data in the region $x< 0.01$, see \cite{Hentschinski:2012kr,Hentschinski:2013id} for a fit with the unintegrated gluon evolved with next-to-leading (NLO) Balitsky-Fadin-Kuarev-Lipatov (BFKL) evolution. The same phase space restriction $x<.01$  has been used in the fits  based on leading order running coupling \cite{Albacete:2010sy}   and NLO \cite{Beuf:2020dxl} Balitsky-Kovchegov evolution, which is the non-linear low $x$ evolution equation formulated in terms of the dipole amplitude. It is therefore natural to use $x<0.01$ as an upper limit on phase space region, where the above arguments can be expected to be applicable. It is needless to say that more conservative bounds would yield even stronger justification for the underlying low $x$ approximations. \\

In the recent paper \cite{Hentschinski:2021aux} we provided numerical evidence  for this proposal, see also \cite{Kharzeev:2021yyf} for related work. In this way entropy at the partonic level is determined through 
\begin{align}
    \label{eq:master}
    S_{\text{partonic}}\left(x, Q^2\right) 
    &= \ln \left[ xg\left(x, Q^2\right) + x\Sigma\left(x, Q^2\right)\right],
\end{align}
where $g(x, \mu_f^2)$ denotes the gluon distribution function at the factorization scale $\mu_f$ and $\Sigma(x, \mu_f^2) = \sum_{a=1}^{n_f} \left(q_a(x, \mu^2) +\bar{q}_a(x, \mu^2)  \right)$ the quark flavor singlet distribution, with $q(x, \mu^2)$ and  $\bar{q}(x, \mu^2)$ quark and antiquark distribution functions for flavor $a$; this contribution is absent in \cite{Kharzeev:2017qzs} and  has been added by two of us in \cite{Hentschinski:2021aux}. It was further found that a framework based on the Balitsky-Fadin-Kuraev-Lipatov\cite{Kuraev:1976ge,Kuraev:1977fs,Balitsky:1978ic} (BFKL) formalism and accounting both for quarks and gluons yields satisfactory agreement with data. However, there are remaining puzzles that are still left unanswered.\\

From a formal point of view, the above definition and its identification with the measured hadronic entropy has the obvious shortcoming that it relates an unphysical object, {\it i.e.} scheme dependent  parton distribution functions (PDFs), to an observable, {\it i.e.} hadronic entropy. From the phenomenological side, the  H1 collaboration determines hadronic entropy from the multiplicities of charged  charged hadrons \cite{H1:2020zpd}. On the other hand,  the theoretical framework of \cite{Kharzeev:2017qzs} is  based on the treatment of purely gluonic emissions which give naturally rise to both charged and neutral hadrons. Comparing therefore the prediction of  \cite{Kharzeev:2017qzs} with  data for charged hadron multiplicities, one  clearly expects a certain mismatch. A related issue is the large discrepancy between partonic and hadronic entropy encountered in \cite{H1:2020zpd}, if the former is evaluated using leading order HERA PDFs. \\

The outline of this paper is as follows: in Sec.~\ref{sec:theory} we provide an overview of ambiguities and open questions in the current description and how they might reflect themselves in the comparison to data. In Sec.~\ref{sec:num} we provide updated numerical results for the BFKL description of experimental results obtained by the H1 collaboration as well as a description based on leading order HERA PDF and PDFs obtained from an unintegrated gluon distribution subject to rcBK evolution. In Sec.~\ref{sec:sat} we provide a first analysis of the transition of this framework towards a region of phase space dominated by non-linear QCD dynamics. In Sec.~\ref{sec:concl} we give our conclusions, while the appendix Sec.~\ref{sec:HSS} summarizes details on the HSS and rcBK unintegrated gluon distribution.

\section{Ambiguities in the current description}
\label{sec:theory}

\subsection{Overall normalization}
\label{sec:norm}

The derivation of Eq.~\eqref{eq:master} 
is based on the expansion of the von Neumann entropy for large $y = \ln 1/x$, or equivalently low $x$ limit which is dominated by gluonic degrees of freedom. One finds
\begin{align}
\label{eq:Spart}
    S_{part.} & = -\sum p_n \ln p_n, &
    p_n (y) & = e^{- \Delta y} \left(1- e^{-\Delta y} \right)^{n-1},
\end{align}
where  $p_n$ denotes the probability to find $n$ dipoles in the proton which
satisfy $p_1(0) = 1$ and $p_{n > 1}(0) = 0$. They are obtained as a solution to the following evolution equation,
\begin{align}
    \label{eq:evolution_prob}
    \frac{d}{d y} p_n(y) & = -\Delta n p_n(y) - \Delta  (n-1) p_{n-1}(y),
\end{align}
with $\Delta$  the BFKL intercept in the 2 dimensional model. 
Even within the 2-dimensional model, the above expression can be slightly generalized to 
\begin{align}
   p_n^{gen.} (y) & = C e^{- \Delta y} \left(1- C e^{-\Delta y} \right)^{n-1},
\end{align}
with a certain constant $C \leq 1$, which yields $p_n (0) \leq 1$ for all $n$. The mean number of dipoles is then obtained as 
\begin{align}
\label{eq:number}
    \langle n \rangle & = \sum_n n p_n = \frac{1}{C} e^{\Delta y} .
\end{align}
Given that the gluon PDF is within the 2 dimensional model subject to the  BFKL equation in zero dimensions,
\begin{align}
    \label{eq:2d_bfkl}
    \frac{d}{d\ln 1/x} xg(x) & = \Delta \cdot xg(x)
\end{align}
and taking into account that PDFs have a certain interpretation as number densities, {\it i.e.} their integral over momentum fraction $x$ yields the expectation value of the parton number operator, it is somehow natural to interpret Eq.~\eqref{eq:number} as the gluon distribution. Even though $xg(x)$ denotes usually the momentum fraction carried by  gluons, while the number density is associated with the integral of $g(x)$, the above identification is correct, since one really determines the mean value of the number of partons per $\ln (1/x)$, {\it i.e.} $\langle d n/dy \rangle $, see also the more detailed discussion in Sec.~\ref{sec:binning}. \\

There arises however an issue, whenever the identification of Eq.~\eqref{eq:number} as the PDF is lifted from the two dimensional model to  four dimensions. While the PDF in four dimensions provides merely information about the one dimensional spatial structure of the proton, it carries an additional dependence on the factorization scale, which can be traced back to an integration over the two remaining transverse dimensions. It is  this additional dependence on the factorization scale as well as the associated scheme dependence of PDFs which prohibit a direct identification of PDFs as number densities. Indeed, for renormalizable gauge theories the number density interpretation applies only  to certain sum rules, {\it i.e.} the difference between the number of quarks and anti-quarks of a certain flavor, which is scheme independent, see {\it e.g.} \cite{Collins:2011zzd} for a detailed discussion. For the gluon distribution such an interpretation may at best hold within {\it e.g.} light-cone gauge at leading order; beyond leading order, the PDF turns scheme dependent. In the light of such complication the best one can hope for is that Eq.~\eqref{eq:number} is proportional to the gluon distribution at leading order, with corresponding modifications, once higher order corrections are invoked. This observation leads us to the slightly modified relation,
\begin{align}
\label{eq:numberMOD}
    \langle n \rangle & = \sum_n n p_n = C^{-1} e^{\Delta Y} = B^{-1} \cdot xg(x), & xg(x) = \frac{B}{C} e^{\Delta y}.
\end{align}
with some so far undetermined parameter $B$. Assuming\footnote{Within high energy factorization, $B=$ const. is a meaningful assumption, since the $x$ dependence is in general determined by the low $x$ resummed gluon density; the same is true for collinear factorization at leading order. Collinear factorization beyond leading order (which is so far not worked out  for this quantity) would most likely yield an $x$ dependent parameter $B$; nevertheless, if the perturbative expansion is converging, this should imply a small correction.} $B=$const. and lifting this expression to four dimensions,  one would then realistically  test in the comparison with data the evolution in $y = \ln 1/x$ and/or in $\ln Q^2$.  Statements about the overall normalization should on the other hand be taken with care. Similarly the quark contribution is absent in the above expression due to the use of the purely gluonic effective 2 dimensional model. Since the low $x$ seaquark distribution is driven by the gluon, the same statement applies to the relevance of this contribution, although it is naturally to be included as long as one interprets $\langle n \rangle$ as the mean number of partons. Sticking for the moment with the gluon distribution\footnote{To include quarks one would merely replace $xg \to xg + x\Sigma$}, the above expression allows to re-express probabilities as
\begin{align}
    \label{eq:pnGen}
    p_n^{gen.} (x) & = \frac{B}{xg(x)} \left(1-\frac{B}{xg(x)} \right)^{n-1},
\end{align}
which finally yields the following partonic entropy,
\begin{align}
    \label{eq:complete_entropy}
 S_{part.} (x) & =   \ln\left[\frac{xg(x) }{B} -1\right] - \frac{xg(x) }{B} \ln\left[1- \frac{B} {xg(x) }\right].
\end{align}
Note that this expression is only meaningful for $xg(x)/B > 1$, which is the region where Eq.~\eqref{eq:pnGen} yields probabilities $p_n \leq 1$. Expanding Eq.~\eqref{eq:complete_entropy} for $xg(x) \gg 1$ and assuming $B = {\cal O}(1)$, one finds 
\begin{align}
\label{eq:+1}
    S_{part.}(x) & = \ln\left[\frac{xg(x) }{B}\right] + 1 + {\cal O}\left[\frac{B} {xg(x) } \right] = \ln\left[\frac{xg(x) }{B/e}\right],
\end{align}
with $e \simeq 2.71828$ Euler's number.
While for asymptotically large $xg(x)$, the contributions due to $B, e$ might be ignored,  for realistic gluon distributions in the kinematic region covered by the HERA experiments, $S_{part.} \leq 4 $ and while $xg(x) \gg 1$ is satisfied, $\ln(xg) =  {\cal O}(1)$ and the above finite terms should be in principle kept. 


\subsection{Charged versus total hadron multiplicity}

The above expression allows to determine partonic entropy as a function of the average number of partons in the system. It is then conjectured that the resulting partonic entropy agrees with the hadronic entropy. The latter can be  obtained from the multiplicity of final state hadrons. Since the detection of neutral hadrons is experimentally challenging, this hadronic multiplicity distribution is usually determined for charged hadrons only. With pions the predominantly produced hadron species, and assuming that final state gluons turn with equal probabilities into positively, negatively, and neutral pion states,  one can as a first estimate assume that the total number of produced hadrons  is roughly $3/2$ times the number of charged hadrons observed in experiment. In turn, the number of gluons and possibly seaquarks which yield charged hadrons is approximate  the fraction $2/3$ of the total parton number. This suggests to correct the partonic entropy by a corresponding factor,
\begin{align}
\label{eq:2/3}
    S_{part.}  \to  S_{charged}  =  S_{part.}  + \ln (2/3).
\end{align}
Clearly this factor is not exact, but merely an estimate of order of magnitude. 

\subsection{Gluon versus Quark contribution}
\label{sec:gvsq}

In \cite{Kharzeev:2017qzs} entanglement entropy has been determined from a 2 dimensional model calculation, based on the dipole picture, which has been related to purely gluonic degrees of freedom. In a follow up study the same authors proposed in \cite{Kharzeev:2021yyf} to evaluate for the kinematic region covered by the HERA experiment the partonic entropy as the logarithm of the seaquark distribution. Finally, in \cite{Hentschinski:2021aux}, two of us proposed to evaluate partonic entropy as the logarithm of the sum of gluon and quark contribution. This treatment was motivated by the observation that the 2 dimensional gluonic model of \cite{Kharzeev:2017qzs}  which identifies entropy as the logarithm of  the average gluon number; for the complete theory, it seems therefore  appropriate to generalize the average gluon number to the average number of quarks and gluons combined. While \cite{Hentschinski:2021aux} managed to successfully describe HERA data, this description was plagued by several shortcomings. While the description based on the HSS gluon of \cite{Hentschinski:2021aux} uses an inconsistent combination of overall normalization constants (which we correct in the subsequent numerical study, see also Sec.~\ref{sec:HSS} for a detailed discussion), the description based on NNLO PDFs is subject to the above mentioned scheme dependence of parton distribution functions which allows for $B \neq 1$, as already point out in \cite{Hentschinski:2021aux}. \\

In the numerical study presented in Sec.~\ref{sec:num}, we show that a description based on a combination of quark and gluon contribution yields a  good description of data within uncertainties if Eq.~\eqref{eq:2/3} is being employed, while we set $B=e$, {\it i.e.} we ignore all pre-asymptotic factors  for the time being. While the observed agreement is pleasing, we believe that in the light of the above uncertainties in normalization, the main goal of the following study is the correct description of $x$ and $Q^2$ dependence. The latter are directly related to evolution equations, to which the underlying distributions are subject to.

\subsection{Binning and  comparison to data}
\label{sec:binning}
\begin{figure}[t]
    \centering
    \includegraphics[width=0.95\textwidth]{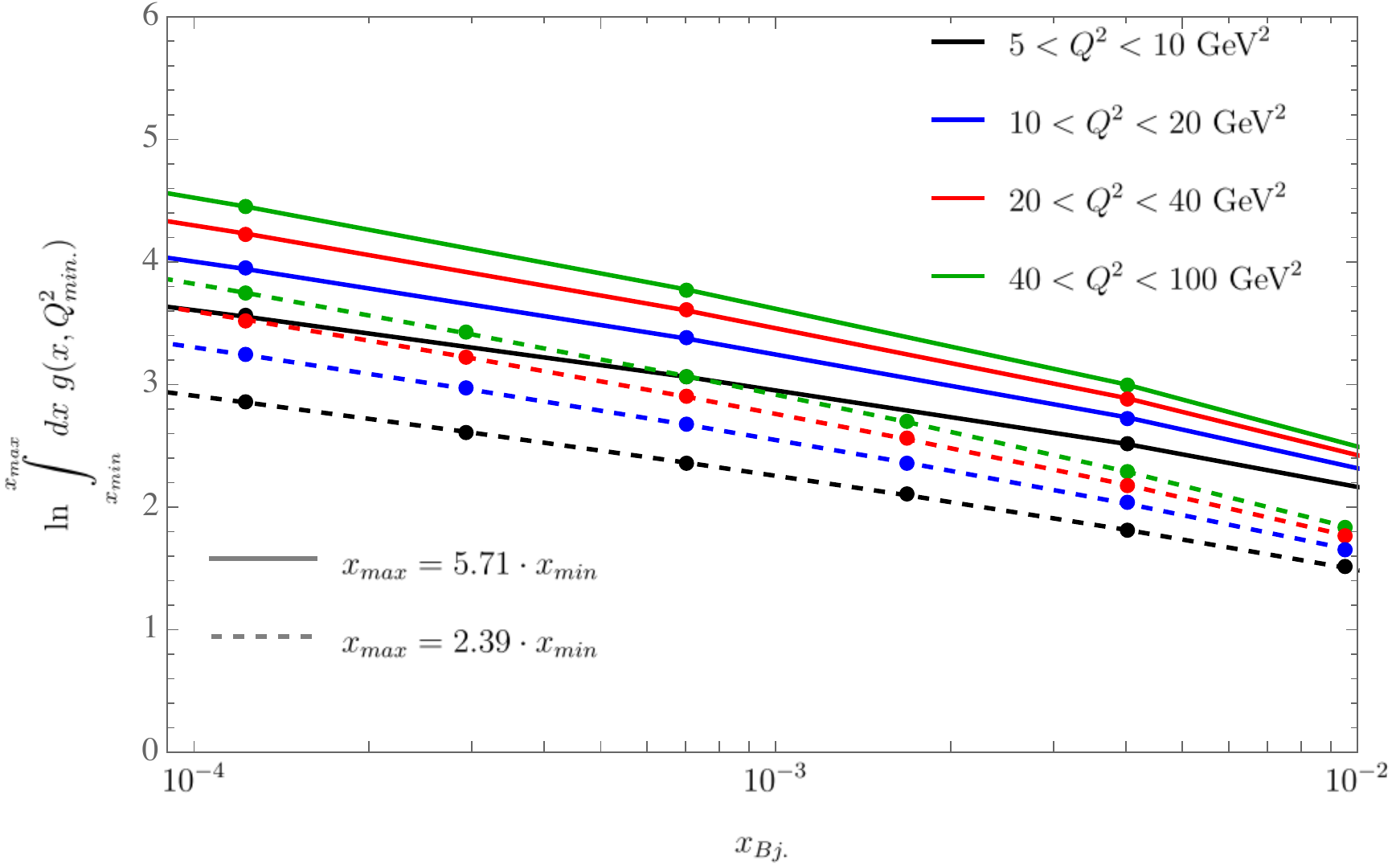}
    \caption{Partonic entropy evaluated  as the logarithm of the number of gluons in a certain bin $[x_{min}, x_{max}]$, see also Eq.~\eqref{eq:partonNumber1} for two different bin sizes. The result clearly depends on the size of the bin. The LO HERAPDF gluon distribution is evaluated at a factorization scale corresponding to the lower $Q^2$ value of the $Q^2$ bin as in \cite{H1:2020zpd}}
    \label{fig:Logpartonnumber}
\end{figure}

Assuming $B=e$ for the moment,  and assuming $\ln(xg) \gg 1$, partonic entropy is directly determined as the logarithm of the number of partons; indeed this has been the original the proposal of \cite{Kharzeev:2017qzs}. In practice this  can however lead to confusion, since parton distribution functions are number densities, {\it i.e.} leaving aside the above mentioned issues of scheme dependence of such distributions, one expects that the number of gluons in a proton is obtained through
\begin{align}
    \label{eq:partonNumber1}
  n_{g}(Q^2) &= \int_0^1 dx \, g(x, Q^2),
\end{align}
with a similar expression in the case of quarks. In DIS reactions, the proton is on the other hand probed at a certain fixed value of Bjorken $x$. Experimentally this corresponds to counting final states for a certain bin size $x \in [x_{min}, x_{max}]$. It is therefore tempting to define the number of gluons at a certain value of $x$ through
\begin{align}
    \label{eq:parton_binning1}
     n_{g}(\bar{x}) &= \int_{x_{\text{min}}}^{x_\text{max}} dx g(x, Q^2),
\end{align}
where $\bar{x} \in [x_{\text{min}}, x_{\text{max}}]$ and might be defined through
\begin{align}
    \bar{x} & = \frac{\int_{x_{\text{min}}}^{x_\text{max}} dx \,x g(x, Q^2)}{\int_{x_{\text{min}}}^{x_\text{max}} dx g(x, Q^2)},
\end{align}
or alternatively as the arithmetic mean of $x_{\text{min}}$ and $x_\text{max}$. Partonic entropy is then obtained as $\ln n_{g}(\bar{x})$. To the best of our understanding, this is the method used by the H1 collaboration to determine the partonic entropy, with $x_{max} \simeq 5.71 x_{min}$, see also the determination of this quantity using this method in Fig.~\ref{fig:Logpartonnumber}. 
While the method yields the number of partons in the region of phase space $[x_{\text{min}}, x_\text{max}]$ if the PDF is interpreted as a number density, the result obviously depends strongly on the size of the interval $[x_{\text{min}}, x_\text{max}]$. In particular the number of partons would approach zero, whenever the size of the interval turns infinitesimal small. For a meaningful comparison it is therefore necessary to study the number of partons, normalized to the bin size. Taking into account that usually one uses logarithmic bins in $x$, {\it i.e.} binning takes place in $y = \ln(1/x)$, one arrives at 
\begin{align}
    \label{eq:norm_av}
    \bar{n}_g(\bar{x}) & = 
    \frac{1}{y_{max} - y_{min}} \int_{y_{min}}^{y_{max}} dy \frac{dn_g}{dy} =
    \frac{n_g(y_{max}) - n_g(y_{min})}{y_{max} - y_{min}},
   & y_{max, min} = \ln 1/x_{min, max},
\end{align}
which in the limit of infinitesimal small bin sizes leads to 
\begin{align}
    \label{eq:infin_bin}
    \bar{n}_g(x, Q^2) & = \frac{d n_g}{d \ln(1/x)} = xg(x, Q^2).
\end{align}
Despite of the usual association of $xg(x, Q^2)$ with momentum sum rules, this quantity is therefore indeed the correct expression to compare to data binned in $Q^2$ and $Y = \ln(1/x)$. For the following numerical study, we further add the contribution due to quarks and  average over the bin size in $Q^2$,
\begin{align}
     \label{eq:binsizeeq}
    \langle  \bar{n}(x, Q^2)  \rangle_{Q^2} & = \frac{1}{Q_\text{max}^2 - Q_\text{min}^2} \int_{Q^2_\text{min}}^{Q^2_\text{max}} d Q^2 \left[ xg(x, Q^2) + x\Sigma(x, Q^2) \right],
\end{align}
which finally yields the expression used for our comparison to data
\begin{align}
    \label{eq:master_entropyPartons}
    \langle  S(x, Q^2)\rangle_{Q^2} & =  \ln  \langle  \bar{n}(x, Q^2)  \rangle_{Q^2}.
\end{align}

\section{Numerical results}
\label{sec:num}

Despite the above ambiguities in the overall normalization, we believe that is meaningful to compare at the current level of accuracy theory predictions to data. Ambiguities are mainly due overall normalization constants which for the partonic entropy turn into additive constants. In the following we use three theoretical models (to be described in more detail in the Appendix):
\begin{itemize}
    \item Leading order HERA PDF, subject to DGLAP evolution \cite{H1:2015ubc}.
    This particular set was chosen due to the observed mismatch between partonic and hadronic entropy by the H1 collaboration (which use LO HERAPDF). The goal of the comparison to LO HERAPDF is to demonstrate that such comparison should be done in the way outlined in the previous section, {\it i.e.} in bins of rapidity.
    \item Leading order PDFs calculated from the HSS unintegrated gluon, subject to BFKL evolution. The gluon density obtained in this scheme accounts for NLO corrections to the  evolution kernel together with a collinear resummation of enhanced NLO contributions. The higher order corrections slow down the growth of the gluon density in the low $x$ region, but do not lead to saturation. The fit is limited to the region $Q^2>2$~GeV$^2$, see \cite{Hentschinski:2012kr,Hentschinski:2013id} for details.   
    \item Leading order PDFs calculated from an unintegrated gluon, subject to rcBK evolution. The gluon density in this framework is subject to leading order BK evolution in $x$;  albeit formally leading order, the evolution kernel includes NLO resummed running coupling  corrections. The evolution takes into account effects due to high gluon densities and leads to a saturated gluon density (for developments that take into account exact kinematics  in combination with saturation effects see \cite{Kutak:2003bd,Kutak:2012rf,Ducloue:2019ezk}). A description of HERA data for the proton structure function $F_2$ is possible, since the nonlinear term in the evolution tames the rapid growth of the gluon distribution, induced by the linear term. 
\end{itemize}
The first  set has been used in \cite{H1:2020zpd} to compare to data. We show that once the corrective factor Eq.~\eqref{eq:2/3} is taken into account and the number of partons is evaluated as discussed in Sec.~\ref{sec:binning}, this set of PDFs gives actually a good description of data, in contrast to the observation made in \cite{H1:2020zpd}, leaving aside a small off-set in the normalization, which might be traced back to the effects discussed in Sec.~\ref{sec:norm}.  The second set has been used previously in \cite{Hentschinski:2021aux}, while the last set is obtained from an unintegrated gluon distribution subject to Balitsky-Kovchegov (BK) evolution \cite{Balitsky:1995ub,Kovchegov:1999yj} which allows us to investigate possible contributions due to non-linear QCD evolution. In all three cases we use Eq.~\eqref{eq:master} to compare to data. The contribution due to seaquarks is in general small; the normalization of the description based on leading order HERA PDFs would slightly improve if one would only consider the contribution due to the gluon. \\

\begin{figure}[t]
    \centering
    \parbox{0.49\textwidth}{\includegraphics[width=0.49\textwidth]{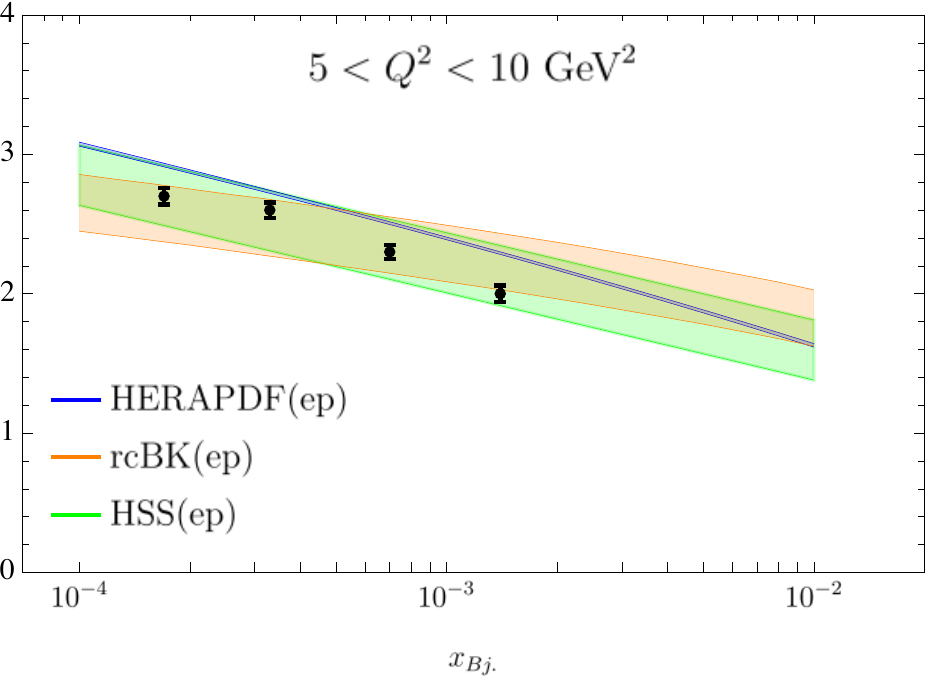}}
    \parbox{0.49\textwidth}{\includegraphics[width=0.49\textwidth]{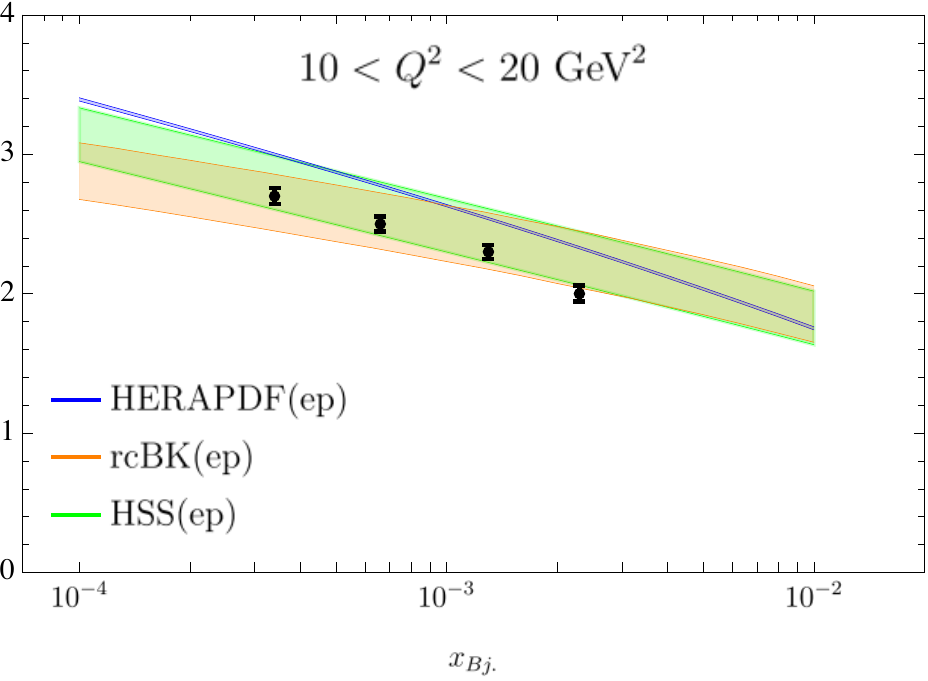}}
     \parbox{0.49\textwidth}{\includegraphics[width=0.49\textwidth]{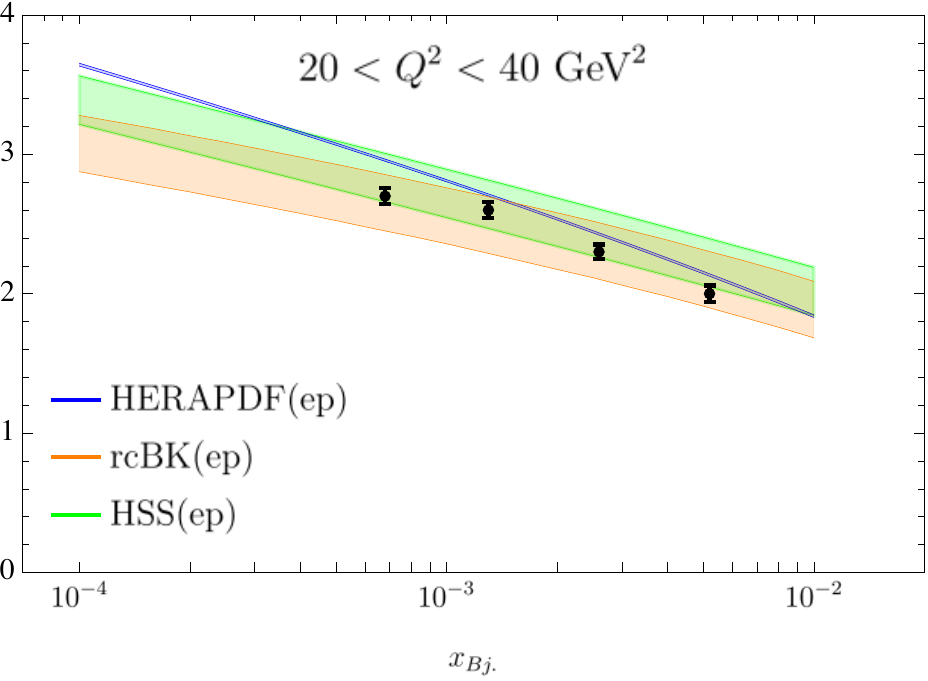}}
      \parbox{0.49\textwidth}{\includegraphics[width=0.49\textwidth]{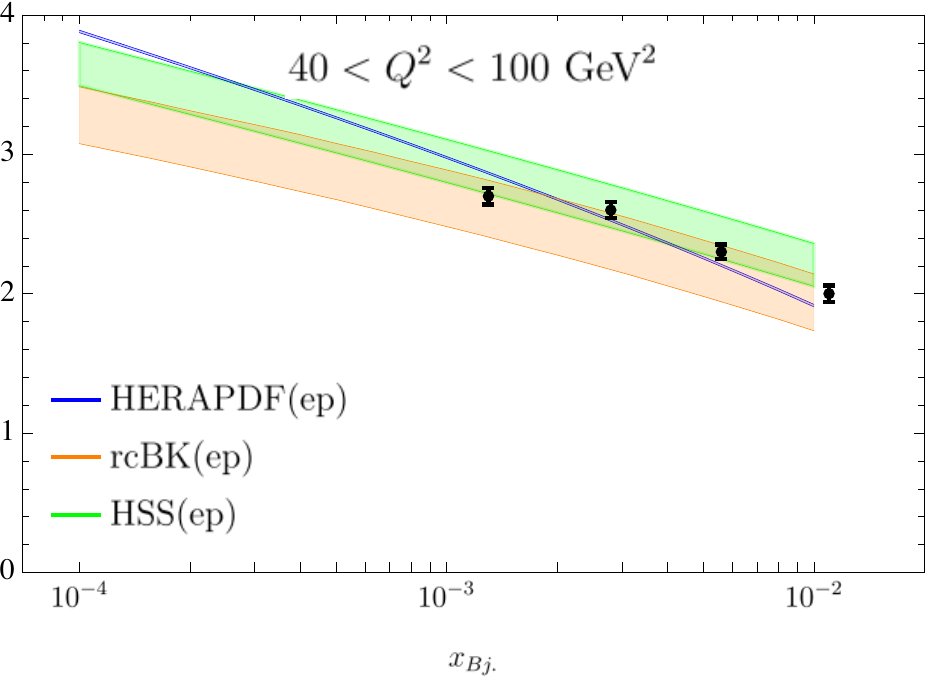}}
    \caption{Partonic entropy corrected for charged hadrons only $\ln \left(xg + x\Sigma \right) + \ln (2/3)$ versus Bjorken $x$.  Results are compared to the final state hadron entropy derived from the charged multiplicity distributions measured by the H1 collaboration \cite{H1:2020zpd} for track pseudorapidities $\eta^*$ in the hadronic centre-of-mass frame  restricted to the range $0< \eta^* < 4$. }
    \label{fig:results}
\end{figure}
To  obtain entropy from the low $x$  QCD  evolution equations,  we need to determine PDFs {\it i.e.} integrated parton densities.
The gluon density is obtained from
\begin{align}
  \label{eq:gluon_collinear}
  xg(x, \mu_F) & = \int_0^{\mu_F^2} d \kt^2 {\mathcal{F}}(x, \kt^2),
\end{align}
where ${\cal  F}(x,\kt^2)$  is the unintegrated gluon distribution, which is obtained from a solution to  BFKL or
BK evolution equations, including a fit to data. To obtain the  quark PDF we apply the Catani-Hautmann procedure
\cite{Catani:1994sq}
\begin{align}
  \label{eq:XSea}
  x\Sigma(x, Q)& = \int_0^\infty \frac{d \Dt^2}{\Dt^2} \int_0^\infty d\kt^2 \int_0^1 dz \Theta\left(Q^2 - \frac{\Dt^2}{1-z} - z \kt^2  \right) \tilde{P}_{qg}\left(z,\frac{\kt^2}{\Dt^2} \right) \mathcal{F}(x, \kt^2),
\end{align}
where the splitting function reads \cite{Catani:1994sq}
\begin{align}
  \label{eq:splitting}
\tilde{P}_{qg}\left(z,\frac{\kt^2}{\Dt^2} \right) & = \frac{\alpha_s 2n_f}{2 \pi} T_F \frac{\Dt^2}{[\Dt^2 + z(1-z) \kt^2]^2} \left[z^2 + (1-z)^2 + 4 z^2 (1-z)^2 \frac{\kt^2}{\Dt^2} \right],
\end{align}
and  $\mu_F$ denotes the factorization scale which we identify for the current study with the photon virtuality $Q$. $\kt$ denotes the gluon transverse momentum and $\Dt = \qt -z\kt$ with $\qt$ the $t$-channel quark transverse momentum; $T_F=1/2$.
$ {P}_{qg}(z)  =  \frac{\alpha_s 2 n_f}{2 \pi} T_F \left[z^2 + (1-z)^2 \right] $.
  In \cite{Hentschinski:2021aux} the HSS integrated gluon Eq.~\eqref{eq:gluon_collinear} has been determined using a version of the unintegrated gluon with an overall normalization inconsistent with the normalization used for the determination of the quark PDF. In the following numerical study this has been corrected. In more general terms, the possibility of such inconsistencies can be traced back to a relatively large uncertainty in the overall normalization of the unintegrated gluon distributions, which have been obtained through fits which rely on the use of the leading order virtual photon impact factor. Since the latter is -- as the seaquark distribution -- proportional to $\alpha_s$, it induces a relatively large normalization uncertainty on the extracted unintegrated gluon. To assess these uncertainties we multiply in the case of the HSS distributions  the integrated gluon distribution by a factor $\alpha_s(Q^2)/\alpha_s(\mu^2)$ and vary the renormalization scale $\mu$ in the range  $\mu \in [Q/2, 2\cdot Q]$. The unintegrated dipole gluon density subject to  rcBK  evolution is obtained through
\begin{equation}
    {\cal F}(x,k^2)=\frac{N_c k^2 S_\perp}{8\pi^2\alpha_s}\int d r^2(1- N(r,x))J_0(r^2k^2),
\end{equation}
where $N (r,x)$ is dipole amplitude obeying BK  equation in coordinate space, see  App.~Sec.~\ref{sec:HSS} for more details. In the above expression $\alpha_s$ is kept constant, while it is running in the kernel of the evolution equation. In order to obtain uncertainties we use $\alpha_s \in [0.2, 0.3]$, which are typical values for the hard scales investigated in this study. For the description based on leading order HERA PDFs  we show the relatively small leading order PDF uncertainties, making use for the numerical evaluation of the package \cite{Clark:2016jgm}.\\

We find in Fig.~\ref{fig:results} that the partonic entropy Eq.~\eqref{eq:master_entropyPartons} obtained from the total number of partons gives a still a very good description of  H1 data  \cite{H1:2020zpd}.
In particular within uncertainty bands, both HSS and rcBK  give a good description of  data. The HERA PDFs slightly overshoot the data but not as drastically as presented in \cite{H1:2020zpd}. This is due to the factor $\ln(2/3)$ which corrects for the fact that only charged hadrons have been measured as well as the effects discussed in Sec.~\ref{sec:binning}. Since gluon and quark contribution add up, the description would slightly improve without the latter.

\section{Towards the real photon limit}
\label{sec:sat}

\begin{figure}[t]
    \centering
    \parbox{0.49\textwidth}{\includegraphics[width=0.49\textwidth]{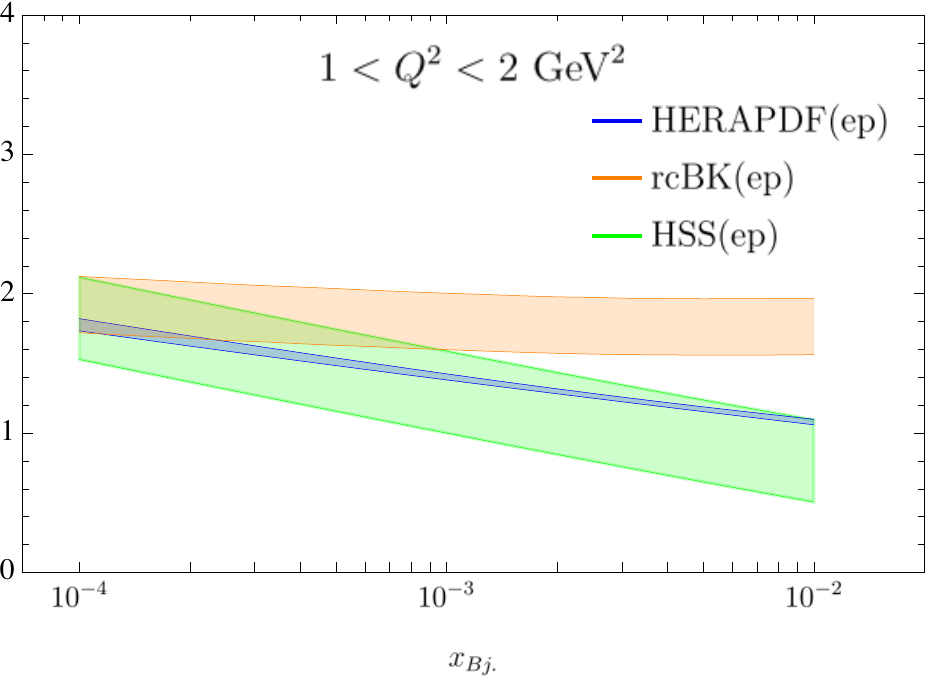}}$\,$ \parbox{0.49\textwidth}{\includegraphics[width=0.49\textwidth]{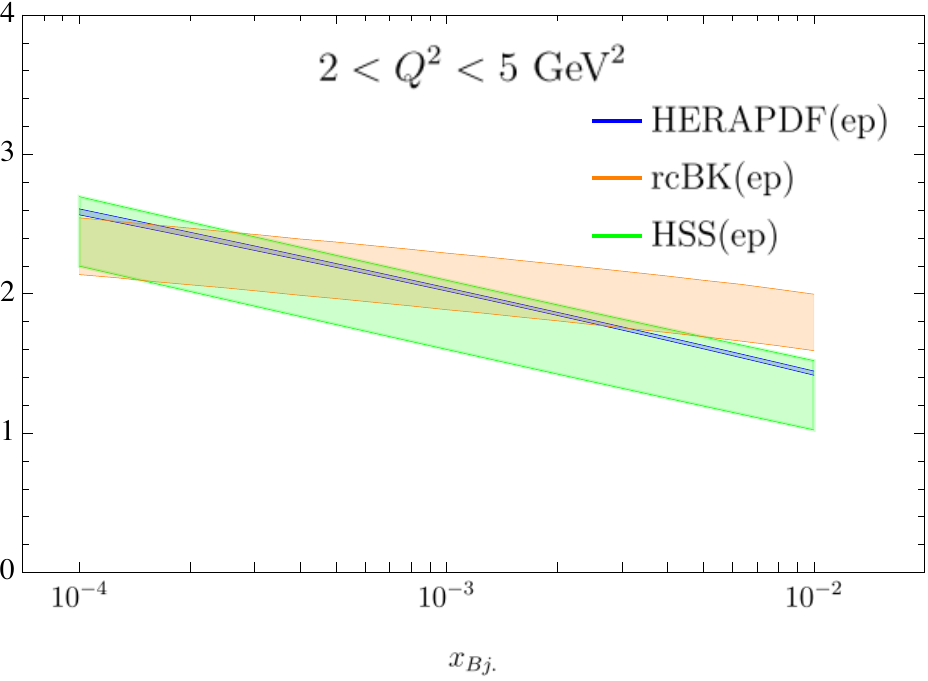}}
    \caption{Partonic entropy  corrected for charged hadrons only, $\ln \left(xg + x\Sigma \right) + \ln (2/3)$ versus Bjorken $x$,  calculated for low $Q^2$ bins. The result demonstrates saturation of entropy at low $Q^2$  and low $x$.  }
    \label{fig:results2}
\end{figure}

Having described the  moderate and large $Q^2$ data, a natural  question to ask is what happens if we go to lower values of  $Q^2$. With $1/Q^2$ the effective area resolved in  the interaction of virtual photon and proton, the limit $Q^2 \to 0$ naturally leads to the case where the photon would observe the entire proton; entanglement entropy should be therefore absent in this limit. While the complete description of such a scenario is still unknown, we believe that investigating this limit within the current framework is already of interest. A more complete description
should be probably formulated  within the more general frameworks\cite{Kovner:2022jqn,Gelis:2010nm}, where  saturation effects \cite{Gribov:1983ivg} and therefore classicalization of wee gluons are taken into account \cite{Dvali:2021ooc}. For the moment we do not consider such modifications. Instead we will investigate the limit of small photon virtualities $Q^2$.
In  Fig.~\ref{fig:results2} we show entanglement entropy as obtained from the  solution to the rcBK evolution equation, as well as the HSS gluon and leading order HERA PDFs. While the rise with $x$  flattens for  three descriptions, this effect is clearly stronger for the rcBK description, which we attribute to effects related to gluon saturation. 
For leading order HERA PDFs and the HSS gluon, we limit the calculations to values $Q^2>1$~GeV$^2$ since for smaller values these descriptions would break down. This is however  not a limitation for the rcBK or Golec-Biernat Wusthoff saturation model \cite{Golec-Biernat:1998zce}, which has been fitted for smaller values of $Q^2$. From a formal point of view, one may justify this through the presence of a semi-hard saturation scale $Q_s(x) > \Lambda_{\text{QCD}}$.  While results for $Q^2<1$~GeV$^2$ must be interpreted with care, they allow for a first qualitative investigation of this region of phase space. We will proceed with analytical studies of this region  using the GBW model. 
\begin{figure}[t]
    \centering
    \parbox{0.49\textwidth}{\includegraphics[width=0.50\textwidth]{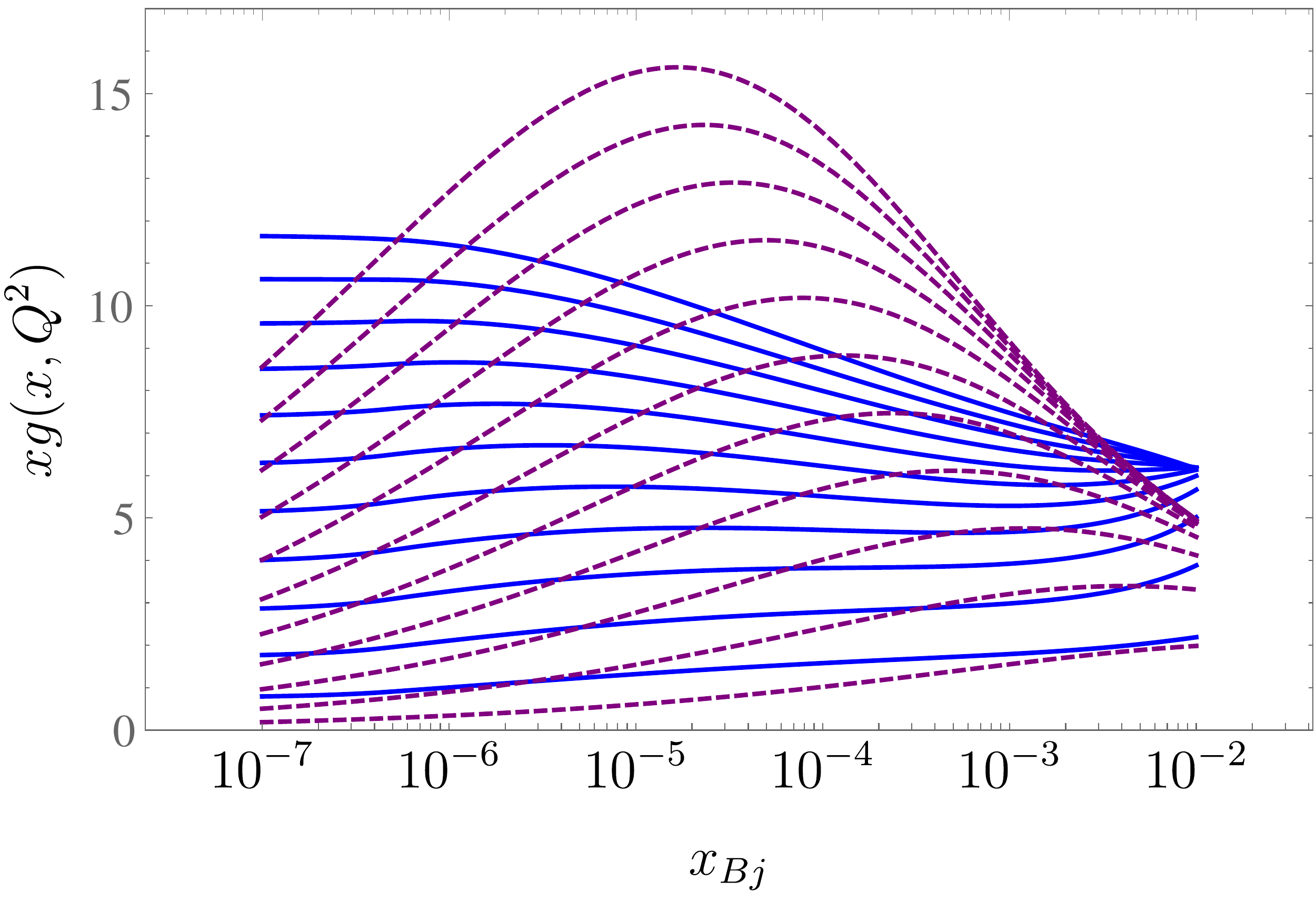}}$\,$ \parbox{0.49\textwidth}{\includegraphics[width=0.50\textwidth]{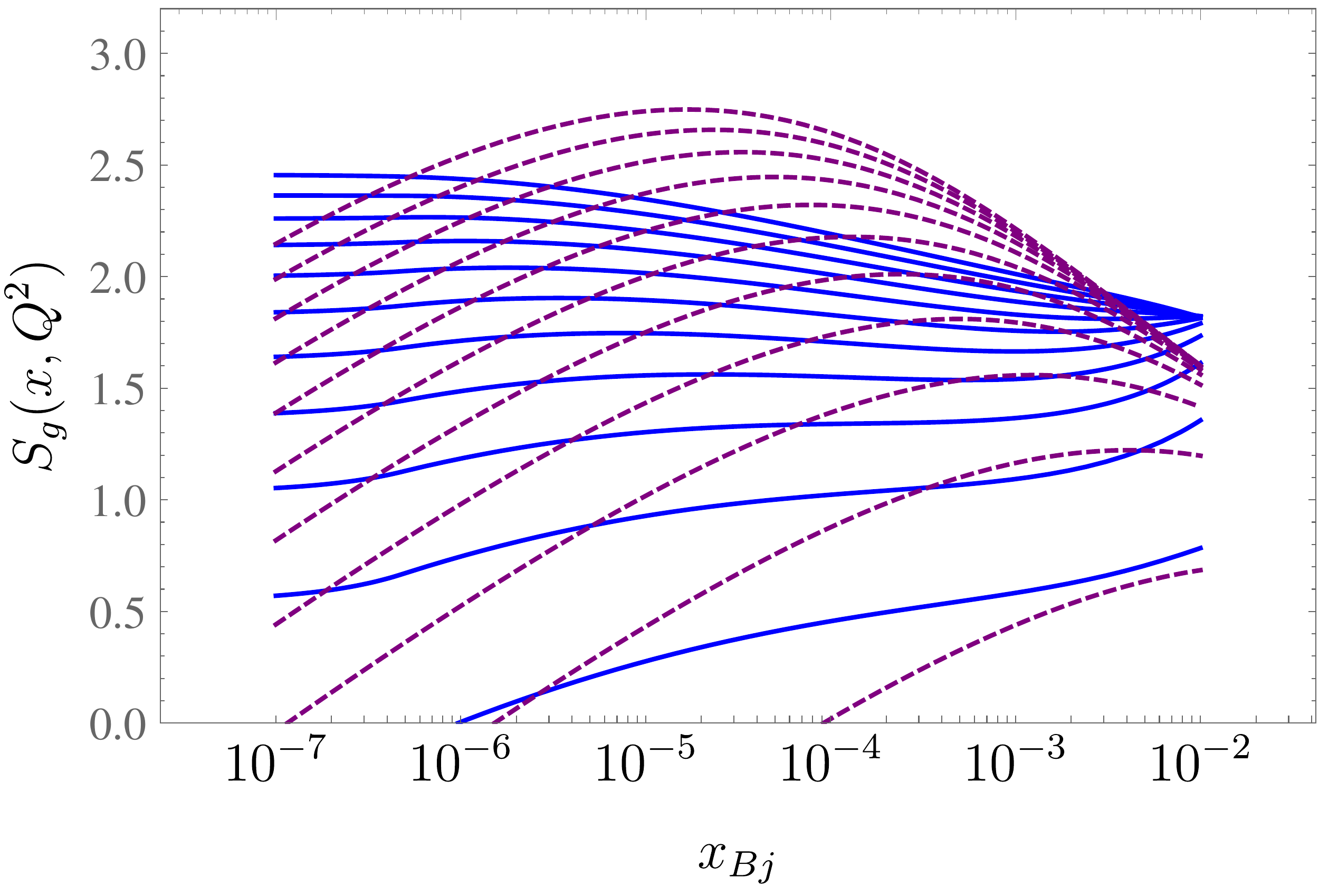}}
    \caption{Gluon PDF (left) and entropy (right)  from rcBK (blue continuous lines) and GBW (purple dotted lines) unintegrated gluon distributions, versus Bjorken $x$, as given by Eq.~\eqref{eq:gluon_collinear} and Eq.~\eqref{eq:master} (for gluons only) calculated for low $Q^2$ values. The result demonstrates saturation and decrease of entropy at low $Q^2$  and low $x$. The values are $Q^2=0.3$~GeV$^2$ (bottom of the plot) up to $2.3$~GeV$^2$ (top of the plot).}
    \label{fig:resultsBK}
\end{figure}
Within this model, the dipole unintegrated gluon density reads\footnote{The BK  for dipole gluon density has very similar $x$ and $k_\perp$ dependence for $|k_\perp|<Q_s$}: 
\begin{equation}
    {\cal F}(x,k^2)=\frac{N_cS_\perp}{\alpha_s8\pi^2}\frac{k^2}{Q_s^2}e^{-k^2/Q_s^2}
\end{equation}
where $Q_s=Q_0^2\left(\frac{x_0}{x}\right)^\lambda$.
The integrated gluon distribution is then obtained as\footnote{we use the 4 flavor fit of \cite{Golec-Biernat:1998zce} with the fit parameter values $S_\perp=14.55$~mb, $x_0=0.41\,10^{-4}$, $\lambda=0.277$, $Q_0^2=1$~GeV$^2$.}
\begin{align}
    \label{eq:gbw_PDF}
    xg(x, Q^2) & = \frac{N_cS_\perp}{\alpha_s8\pi^2} \left[ Q_s(x)^2 \left(1- e^{Q^2/Q_s^2(x)} \right)
    -
    Q^2 e^{Q^2/Q_s^2(x)}
    \right].
\end{align}
For large photon virtualities $Q^2 \gg Q_s(x)^2$ this yields
\begin{align}
    \lim_{Q^2 \gg Q_s^2} xg(x, Q^2) & = \frac{N_cS_\perp}{\alpha_s8\pi^2} Q_s^2,
\end{align}
{\it i.e.} the integrated gluon distribution is directly proportional to the saturation scale. In the limit of small photon virtualities one finds on the other hand: 
\begin{equation}
   \lim_{Q^2 \ll Q_s^2} xg(x, Q^2) =\frac{N_c}{\alpha_s16\pi^2}\frac{S_\perp Q^4}{ Q_s^2(x)} \sim x^{\lambda},
\end{equation}
{\it i.e.} the integrated gluon distribution turns into a  falling function of $x$. In particular one has the formal limit $xg(x, Q^2 = 0) = 0$. For entanglement entropy one finds, extrapolating the current framework to low $Q^2$, 
\begin{align}
    \lim_{Q^2 \ll Q_s^2} 
    S(x,Q^2)=\ln\left(\frac{S_\perp Q^4}{Q_s^2(x)}\right)+ \ln \frac{N_c}{16\alpha_s\pi^2}.
\end{align}
Depending on the precise values of the transverse size $S_\perp$, this expression will for some value of $x$ turn eventually negative. Note however that the definition of probabilities Eq.~\eqref{eq:pnGen} requires $xg(x) \geq B = e$, for the current setup, which prevents us from reaching negative values.   On the other hand for $Q^2 \gg Q_s(x)^2$ we have
\begin{equation}
    \lim_{Q^2 \gg Q_s^2}   S(x,Q^2)=\ln\left(S_\perp Q_s^2(x)\right)+ \ln \frac{N_c}{8\alpha_s\pi^2} = \lambda \ln \frac{1}{x} + \text{const.},
\end{equation}
{\it i.e.} we recover the original expression for partonic entropy obtained in \cite{Kharzeev:2017qzs} plus a certain constant contribution.\\

\begin{figure}[t]
    \centering
     \centering
    \parbox{0.48\textwidth}{\includegraphics[width=0.48\textwidth]{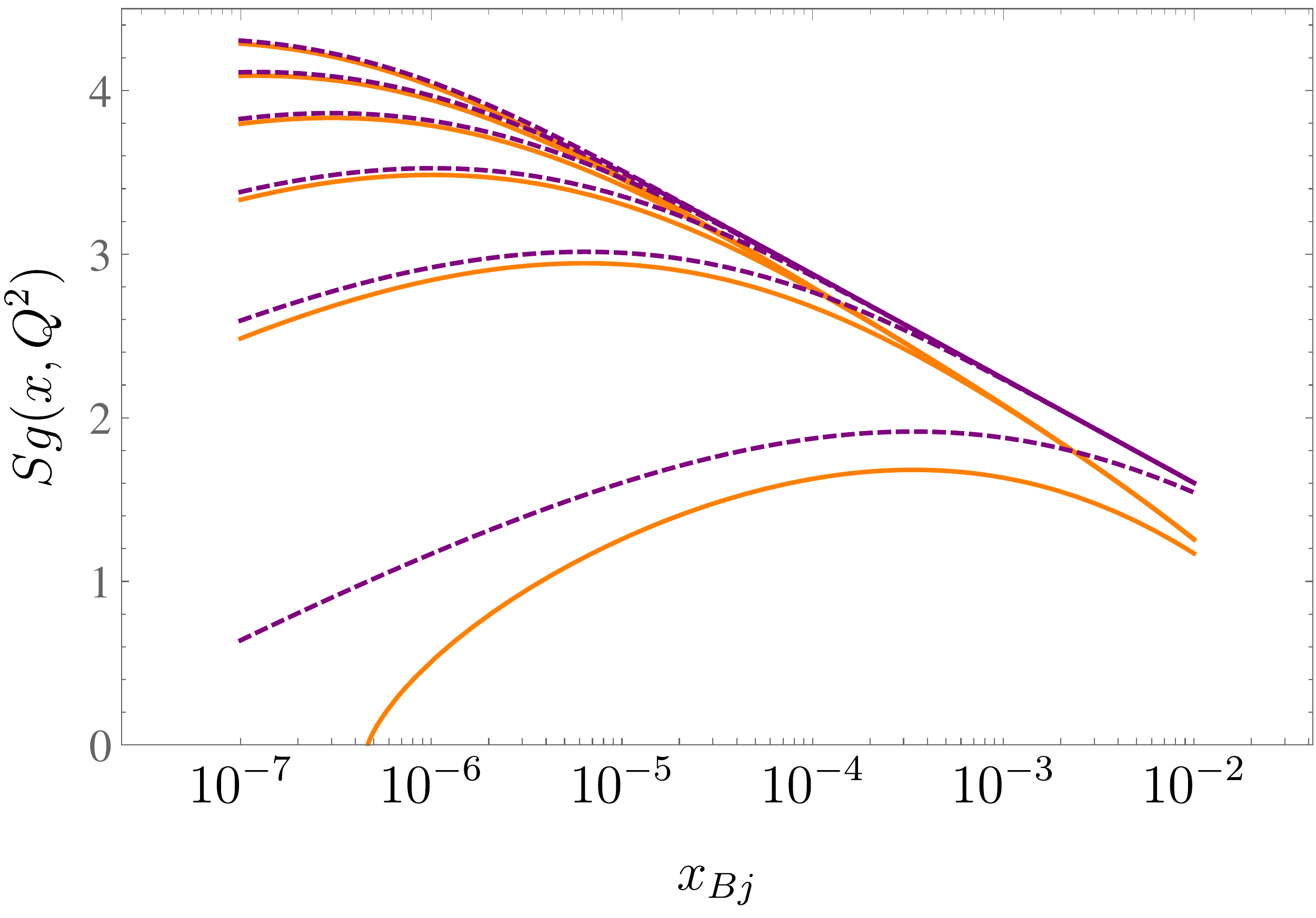}}$\quad$ \parbox{0.48\textwidth}{\includegraphics[width=0.48\textwidth]{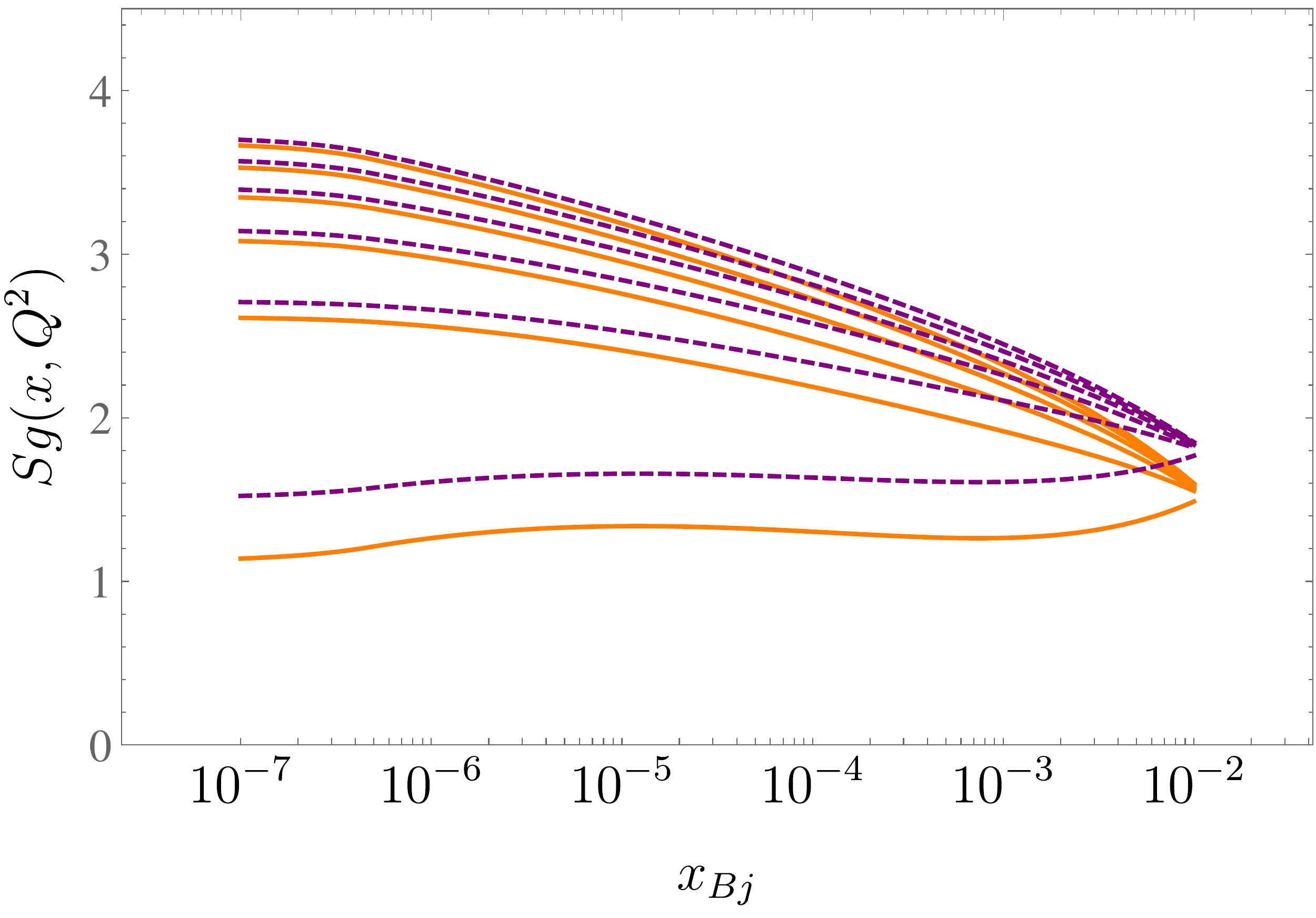}}
    \caption{Left: GBW partonic entropies for increasing $Q^2$ from bottom to top  with $Q^2=1$~GeV$^2$ the lowest and $Q^2=11$~GeV$^2$ the highest value with  increments by $Q^2=2.5$~GeV$^2$, versus Bjorken $x$, as given by Eq.~\eqref{eq:master_entropyPartons} (orange continuous) and Eq.~\eqref{eq:+1} (purple dashed) assuming $B=e$. Right: the same as left but the integrated gluon obtained from the rcBK unintegrated gluon  is used  to determine entropy.}
    \label{fig:ratios}
\end{figure}

A numerical study of the GBW model compared to rcBK is provided in Fig.~\ref{fig:resultsBK}. 
In the figure we plot only the contributions due to gluons. Entropy is being evaluated  at low values of $Q^2 < 1$~GeV$^2$ since the both gluon densities have been fitted for this region of phase space. Moreover there exist in principle data for this region of phase and a comparison to those data would be of high interest, once they are in available in a suitable form. From a formal point of view, the existence of a semi-hard saturation scale allows for the evaluation of the gluon density at such low scales. We however stress here that the entropy formula in this region is an  extrapolation. 
As can be seen from the plot to the right, the integrated gluon distribution is no longer necessarily large at lowest values of $x$. The expansion of Eq.~\eqref{eq:complete_entropy} for large $xg(x)$ is therefore not necessarily a good approximation. We therefore compare in Fig.~\ref{fig:ratios} both the exact formula for entropy, derived within the 2 dimensional model, and its asymptotic expansion. The exact formula is not necessarily more accurate than the asymptotic expression, but the observed deviation indicates in which regions of phase space effects due to non-linear QCD dynamics might become relevant. In all of the the calculations we assume $B=e$ which is the choice which minimizes the contribution due to constant terms. 
Note that the observation that for certain values  of $Q^2$ the gluon PDF is a falling function of $x$ can be directly linked to the behavior of the dipole amplitude featuring saturation stemming from  a solution to BK or JIMWLK  \cite{Jalilian-Marian:1997jhx,Jalilian-Marian:1997qno,Jalilian-Marian:1996mkd,Iancu:2000hn} evolution and which has been already observed for the solution of the BK equation in \cite{Kutak:2006rn}. We interpret this as a mechanism of localization due to saturation of wee partons in longitudinal direction. \\

\section{Conclusions}
\label{sec:concl}

In this paper we continued the study of the proposal formulated in \cite{Kharzeev:2017qzs} to calculate for DIS reactions entanglement entropy from parton distribution functions. The purpose of this study has been twofold: on one hand we attempted to clarify some of the lose ends of this proposal such as the overall normalization, the relation between entanglement entropy  and hadronic entropy of charged hadrons as well as a discussion on how to evaluate the number of partons for the determination of partonic entropy. On the other hand we provided a first exploration of the proposal towards low photon virtualities and very low $x$, where eventually non-linear QCD dynamics is expected to manifest itself.

To this end we provided a  description of the  data measured by the H1 collaboration through parton distribution functions subject to  BFKL and BK evolutions equations as well as leading order HERA PDFs. As outlined in section 2, the description is at the moment mainly qualitative, due to various ambiguities in the theoretical description. In particular a precise phenomenology would require to work out an appropriate factorization of entanglement entropy into parton distribution functions and corresponding perturbative coefficients  which would guarantee factorization scheme independence of the framework beyond leading order.
While the frameworks used in this paper agree for moderate and large values of $Q^2$, we find differences between BFKL evolution and leading order PDFs on the one hand and BK evolution on the other hand, if we turn to smaller values of $Q^2$.  In particular nonlinear dynamics, as encoded in the BK equation, suggests that entropy saturates or even decreases. On technical level we link this behavior to the feature of the BK  gluon PDF which predicts a valence like gluon PDF {\it i.e.} a  distribution falling as a function of $x$ at some low enough value of $x$. This behavior of  entanglement entropy is expected since at very low values of $Q^2$ the photon virtuality is so low that the system can not be bi-partitioned and  entanglement entropy needs to vanish.  To provide a definite answer to those questions, a generalization of the framework discussed here is necessary; in particular to explain the observed decrease of entropy with $x$.


\section*{Acknowledgments}
\label{sec:Ack}
We would like to thank Valerio Bertone, Krzysztof Golec-Biernat, Franceso Hautmann, Alex Kovner, Genya Levin, Misha Lublinsky, Al Mueller,  Jacek Otwinowski, Andreas Sch\"afer, Zhoudunming Tu, Raju Venugopalan for stimulating discussions.\\ 
KK acknowledges the support of The Kosciuszko Foundation for the Academic year 22/23 for the project 
"Entropy of  dense system of quarks and gluons".
MH is grateful for hospitality at the Institute of Nuclear Physics and acknowledges support by
Consejo Nacional de Ciencia y Tecnolog\'ia grant number A1 S-43940
(CONACYT-SEP Ciencias B\'asicas).\\
\appendix

\section{Some details on the rcBK  and HSS  }
\label{sec:HSS}

The procedure of  getting the HSS unintegrated gluon density relies  on solving  NLO BFKL equation with BLM  prescription for scale choice, collinear improvements and fitting initial condition to HERA data \cite{Hentschinski:2012kr,Hentschinski:2013id}. The rcBK is basically the LO BK equation formulated in the coordinate space and with running coupling constant included. 
The initial conditions are fitted  \cite{Albacete:2010sy} to HERA data \cite{H1:2009pze}.

\begin{itemize}
\item
The unintegrated dipole gluon density from the BK is obtained via
\begin{equation}
    {\cal F}(x,k^2)=\frac{N_c k^2 S_\perp}{8\pi^2\alpha_s}\int d r^2(1- N(r,x))J_0(r^2k^2)
\end{equation}
where $N (r,x)$ is dipole amplitude obeying BK  equation in the coordinate space. We use $S_\perp=16.4$~mb and $\alpha_S$ is kept constant\footnote{This is choice often use once  obtaining gluon density form the BK  equation solved in coordinate space (see for example). The coupling constant at LO accuracy cancels for such observable as $F_2$ between impact factor and the gluon density independently on whether it is running (for the same choice  of scale) or whether it is fixed. }. In order to assess normalization uncertainties due to the choice of this coupling constant, we vary it in the range [0.2-0.3]].
The BK equation reads \cite{Balitsky:2006wa}
\begin{align}
    \frac{\pd N(r,Y)}{\pd Y} = \int \d{\vec{r_1}} K^{run}(\vec{r},\vec{r_1},\vec{r_2})
        (N(r_1,Y)+N(r_2,Y)-N(r,Y)- N(r_1,Y)N(r_2,Y))
\end{align}
where, $Y=\ln(x_0/x)$ and  $\vec{r_2}=\vec{r}-\vec{r_1}$.
The kernel $K^{run}$ is given by
\begin{align}
    K^{run}(\vec{r},\vec{r_1},\vec{r_2}) = \frac{\alpha_s(r^2)}{2\pi^2}N_c
    \left[
        \frac{r^2}{r_1^2 r_2^2}
        + \frac{1}{r_1^2}\left( \frac{\alpha_s(r_1^2)}{\alpha_s(r_2^2)}-1\right)
        + \frac{1}{r_2^2}\left( \frac{\alpha_s(r_2^2)}{\alpha_s(r_1^2)}-1\right)
    \right],
    \end{align}
with
\begin{align}
    \alpha_s(r^2) = \frac{12\pi}{(11N_c - 2N_f)\ln\left(\frac{4C^2}{r^2\Lambda_{QCD}^2}\right)};
\end{align}
the maximal allowed value for $\alpha_s$ is $\alpha_{s,max}=0.7$. The initial condition is given by McLerran-Venugopalan model \cite{Albacete:2010sy,McLerran:1993ni}
\begin{align}
    N^{\mathrm{MV}}(r,Y=0)=1-\exp\left[-\frac{(r^2Q_{s0}^2)^\gamma}{4}\ln\left(\frac{1}{r\Lambda_{QCD}} + \mathrm{e} \right) \right]
\end{align}
The equation was solved by newly developed code using Runge-Kutta method and using the parameters in the fit  \cite{Albacete:2010sy} to HERA data \cite{H1:2009pze}. The numerical values of the parameter read 
$Q_{s0}=0.165 GeV$, $\gamma=1.135$, $C=2.52$, $\Lambda_{QCD}=0.241$, $N_f=3$, $S_\perp=16.4$ mb.
\item
The HSS gluon density follows from the  BFKL evolution equation with the kernel obtained at NLL accurancy with collinear
improvements \cite{Hentschinski:2012kr,Hentschinski:2013id, Bautista:2016xnp}.  
The unintegrated gluon density  reads
\begin{align}\label{eq:Gudg}
  \mathcal{F}\left(x, {\bm k}^2, M\right) 
& = 
 \frac{1}{{\bm k}^2}\int \limits_{\frac{1}{2}-i\infty}^{\frac{1}{2} + i \infty}  \frac{d \gamma}{2 \pi i}   \; \; \hat{g}\left(x, \frac{M^2}{Q_0^2}, \gamma \right) \, \, \left(\frac{{\bm k}^2}{Q_0^2} \right)^\gamma
\end{align}
where $M$ is a characteristic hard scale of the process which we identify with $Q$.  $\hat{g}$ is an
operator in $\gamma$ space,
\begin{align}
  \label{eq:23}
 \hat{g}\left(x, \frac{M^2}{Q_0^2}  \gamma \right)
  & = 
 \frac{\mathcal{C}\cdot \Gamma(\delta - \gamma)} {\pi \Gamma(\delta)}  \; \cdot \; 
 \left(\frac{1}{x}\right)^{\chi\left(\gamma \right)} \,\cdot \notag
 \\ 
&  
  \Bigg\{1    + \frac{\bar{\alpha}_s^2\beta_0  \chi_0 \left(\gamma\right) }{8 N_c} \log{\left(\frac{1}{x}\right)} 
  \Bigg[- \psi \left(\delta-\gamma\right)
 +  \log \frac{{M}^2}{Q_0^2} -  \partial_\gamma \Bigg]\Bigg\}\;, 
\end{align}
where $\bar{\alpha}_s = \alpha_s N_c/\pi$ with $N_c$ the number of
colors and   $\chi(\gamma)$ is the next-to-leading
logarithmic (NLL) BFKL kernel which includes a resummation of both collinear enhanced terms as well as  a 
resummation of  large terms proportional to the first coefficient of the QCD
beta function. For the current study we set $M = \overline{M} = Q$ and $n_f=4$ with $\Lambda_{\text{QCD}} = 0.21$~GeV.
$Q_0 = 0.28$~GeV, and $\delta = 6.5$. have been determined from a fit to  the $F_2$ structure function in \cite{Hentschinski:2012kr}. In this fit the  overall running coupling constant  has been evaluated at the renormalization scale $\mu^2=QQ_0$, with $Q$ the photon virtuality. For the construction of parton distribution function $\mu^2=Q^2$  is however a more natural choice. We therefore reevaluated the underlying fit and found that data on the proton structure $F_2$ \cite{H1:2009pze}  are equally well described, if we use  $\mu^2=Q^2$ for the photon impact factor with a  normalization  $\mathcal{C}=4.31$. It is then this convention which we  use in this study. In \cite{Hentschinski:2021aux} the integrated quark distribution has been evaluated using $\mathcal{C}=4.31$, while the  gluon has been evaluated with the normalization constant corresponding to the scale choice $\mu^2=QQ_0$. Since distributions arise from the same unintegrated gluon distribution, such a treatment is not consistent. This has been corrected in the present study, including a variation of the overall running coupling to indicate the normalization uncertainty of our result. 
\end{itemize}

\end{document}